\documentclass[11pt]{article}

\usepackage{graphicx}
\usepackage{amsmath,amssymb}
\usepackage{amsthm}
\usepackage{color}
\usepackage{enumerate}

\newcommand{\onemath}[2] {\begin{equation}\begin{split}#1 \label{#2}\end{split}\end{equation}}
\newcommand{\onemathN}[1] {\begin{equation*}\begin{split}#1 \end{split}\end{equation*}}

\newtheorem{theo}{Theorem}[section]

\newcommand{\diffd}{{\rm d}}

\begin{document}

\begin{center}
{
\Large
An algorithm for calculating $D$-optimal designs for polynomial regression with prior information and its appilications\\
}
\vspace{0.5cm}
Hiroto Sekido \footnote{sekido@amp.i.kyoto-u.ac.jp} \\
{\it Department of Applied Mathematics and Physics, Graduate School of Informatics, Kyoto University, Kyoto 606-8501, Japan.}
\end{center}

\vspace{0.5cm}

\begin{center}
\begin{minipage}{10.5cm}

{
\small
{\bf Abstract}:
Optimal designs are required to make efficient statistical experiments.
D-optimal designs for some models are calculated by using canonical moments.
On the other hand, integrable systems are dynamical systems whose solutions can be written down concretely.
In this paper, polynomial regression models with prior information are discussed.
In order to calculate D-optimal designs for these models,
a useful relationship between canonical moments and discrete integrable systems is used.
By using canonical moments and discrete integrable systems,
an algorithm for calculating D-optimal designs for these models is proposed.
Then some examples of applications of the algorithm are introduced.

{\it Key words}:
D-optimal design, canonical moment, polynomial regression model, discrete integrable system, nonautonomous discrete time Toda equation
}
\end{minipage}
\end{center}

\section{Introduction}

In statistics, design of experiments is a methodology to make efficient experiments.
Optimal designs have been studied by numerous authors in the literature.
Especially D-optimal designs have been investigated by many authors.
One of the approaches for D-optimal designs is to use canonical moments.

D-optimal designs and D${}_s$-optimal designs for polynomial regression models were studied in \cite{as33jtuw}.
In the calculation in \cite{as33jtuw}, an important point is to use canonical moments instead of ordinary moments,
and D-optimal designs can be identified by its canonical moments.
As we see in Section \ref{s:oiauf90a8df}, the objective function is written down in term of canonical moments.
Besides this, D-optimal designs for various models have been calculated in \cite{nvcuisd987j, afygfuhs83, mdoiudj91j8d, mdoiudj91j9d, ajifa9d78rjio, asfa873hfi, fdrtd34ai, asfa831221i, jjdfayt8367, asiufad74, ioasf873j, ajsdfghh86, auifa987, auifa988, auifa989, ddesa3fdaea3}.
For example, D-optimal designs for weighted polynomial regression with a weight function $x^\alpha (1-x)^\beta$ were found explicitly in \cite{auifa988} by using canonical moments.
However, in most cases, an explicit form of the D-optimal design is unknown.
We here consider polynomial regression models with some prior information.
For example, D-optimal designs for this kind of models have been studied in \cite{afygfuhs83, asfa873hfi, jjdfayt8367}.
D-optimal designs for polynomial regression models with only odd (or even) degree terms are calculated in \cite{afygfuhs83}.
Polynomial regression without intercept is considered in \cite{jjdfayt8367}.
Polynomial regression models through origin is considered in \cite{asfa873hfi}.

On the other hand,
the term {\it integrable systems} is used for nonlinear dynamical systems whose solutions can be written down concretely.
For Hamiltonian systems with finite degree of freedom, their integrability is defined in the Liouville-Arnold theorem \cite{aosdf39jjd}.
However, even now, there is no mathematical definition of integrability for nonlinear systems with infinite degree of freedom.
This is why we call an explicitly solvable nonlinear system as an integrable system.
Discrete integrable systems are discrete analogues of integrable systems.
That is, it is well-known that integrable systems can be discretized such that descretized systems are also solvable.

Discrete integrable systems have been applied to numerical analysis.
Typical examples are matrix eigenvalue algorithms \cite{cva09sau,o1h3ituw}, and 
algorithms to compute matrix singular values \cite{daed35dsf}.

An algorithm for calculating D-optimal designs for polynomial regression through a fixed point is proposed in \cite{sekidos1}.
In \cite{sekidos1}, the relationship between canonical moments and discrete integrable systems are used.
In this thesis, we propose an algorithm for constructing D-optimal designs for polynomial regression with prior information, which is a generalized algorithm of \cite{sekidos1},
and the considered models include polynomial regression through origin \cite{asfa873hfi}, and some weighted polynomial regression as particular cases.
Moreover the algorithm can be applied to some optimal designs, and it allows us to calculate a larger class of optimal designs.
That means the relationship between canonical moments and discrete integrable systems expands the class of D-optimal designs which can be calculated.

\section{A preliminary} \label{s:oiauf90a8df}

This section give a brief introduction of polynomial regression models, D-optimal designs, and canonical moments.

At first, we consider the following common linear regression model
\onemathN{
& Y = \theta^{\rm T} f(x) + \varepsilon,\\
& E[\varepsilon] = 0, \quad V[\varepsilon] = \sigma^2,
}
where $f(x) = (f_0(x) \; f_1(x) \; \cdots \; f_{m-1}(x))^{\rm T}$ denotes a known vector of linear independent functions,
$\theta = (\theta_0(x) \; \theta_1(x) \; \cdots \; \theta_{m-1}(x))^{\rm T}$ denotes an unknown vector of parameters,
$\varepsilon$ denotes the error term.
Here we assume each experiment has a stastically indendent error term $\varepsilon$.

Let $\mathcal P_I$ be the set of probability mesures on the borel set on $I$.
For given $\mu \in \mathcal P_I$, let $c_k$ denote the $k$th moment $\int_I x^k {\rm d} \mu(x)$.
Suppose the number of expriments be $n$, and experimental conditions $x_1, x_2, \ldots, x_n$ should be in interval $[0, 1]$.
Here we consider design $x_1, x_2, \ldots, x_n$ is corresponding to the probability mesure 
$\mu \in \mathcal P_{[0,1]}$ such that $\mu(\{x\}) = \#\{k|x_k=x\} / n$.
Then D-optimal designs are defined as probability mesure $\mu \in \mathcal P_{[0,1]}$ 
which maximizes the determinant of Fisher information matrix $M_f(\mu) = \int_0^1 f(x) f(x)^{\rm T} {\rm d} \mu (x)$.
In the case of $(m-1)$th polynomial regression, that is, in the case where $f_k(x) = x^k$, 
the D-optimal design is defined as the optimization solution of the optimization problem
\onemath{
& \mbox{maximize}\; |c_{i+j}|_{i,j=0}^{m-1} \\
& \mbox{subject to}\; \mu \in \mathcal P_{[0,1]}.
}{opthoefho}
Note that, in the optimization problem, while $\mu(\{x\})$ should be multiple of $1/n$ for all $x$,
we do not take a such constraint.
Therefore we consider only the relaxation optimization problem.
See \cite{ifa34jy4u} for details.

In the optimization problem \eqref{opthoefho}, the objetive function is written down in terms of moments.
However the constraint is complicated in terms of moments.
To simplify the constraint, sometimes the optimization problem is rephrased in terms of canonical moments.

Now, we define canonical moments.
Suppose we consider the set $\mathcal P_{[0,1]}$ of the probability measures on $[0,1]$.
For a given probability measure $\mu \in \mathcal P_{[0,1]}$, let $c_k^+$ denote the maximum of the
$k$-th moment over the set of all measure having moments $c_0,c_1,\ldots,c_{k-1}$.
Similarly let $c_k^-$ denote the corresponding minimum.
Canonical moments are defined by
\onemath{ p_k = \frac{c_k - c_k^-}{c_k^+ - c_k^-}, \quad k=1,2,\ldots,N, }{eq:adfi8733}
where $N$ is the minimum of $j$ which satisfy $c_{j+1}^+ = c_{j+1}^-$.
If $c_j^+ > c_j^-$ for an arbitrary positive integer $j$, let's set $N = \infty$.

Canonical moments have the property that 
\onemath{
  \begin{cases}
    p_k \in (0,1), & k=1,2,\ldots,N-1, \\
    p_N \in \{0,1\}. & 
  \end{cases}
}{eq:asfuio7h}
Conversely, for a given arbitrary sequence $\{p_k\}_{k=1}^N$ which satisfies \eqref{eq:asfuio7h},
there is a unique $\mu \in \mathcal P_{[0,1]}$ which has the canonical moments $\{p_k\}_{k=1}^N$.

It is to be noted that canonical moments have a Hankel determinant expression.
Let $H_k^{(n)}, \overline H_k^{(n)}$ be the Hankel determinants defined by
\onemath{
  H_k^{(n)} = \left|c_{i+j+n}\right|_{i,j=0}^{k-1} , \quad 
  \overline H_k^{(n)} = \left|c_{i+j+n} - c_{i+j+n+1}\right|_{i,j=0}^{k-1},
}{eq:ksajsdfoiuaf874}
where $k=1,2,\ldots, \;\; n=0,1,2,\ldots$.
We here set that $H_0^{(n)}=1$, and that $H_k^{(n)} = 0$ if a matrix size $k$ is negative.
The canonical moments $p_k$ have the following Hankel determinant expression:
\onemathN{
  p_{2k-1} = \frac{ H_{k}^{(1)} \overline H_{k-1}^{(0)} }{ H_{k}^{(0)} \overline H_{k-1}^{(1)} }, \quad 
  p_{2k} = \frac{ H_{k+1}^{(0)} \overline H_{k-1}^{(1)} }{ H_{k}^{(1)} \overline H_{k}^{(0)} }, \quad k=1,2,\ldots.
}
We introduce useful variable $\zeta_k$ by the transformation of canonical moments $p_k$
\onemath{
  & \zeta_0 = 0, \quad \zeta_1 = p_1, \\
  & \zeta_k = (1-p_{k-1})p_k, \quad k=2,3,\ldots,N. 
}{eq:iofaou4a}
Then $\zeta_k$ also have the Hankel determinant expression
\onemathN{
  \zeta_{2k-1} = \frac{ H_{k}^{(1)} H_{k-1}^{(0)} }{ H_{k}^{(0)} H_{k-1}^{(1)} },\quad
  \zeta_{2k}   = \frac{ H_{k+1}^{(0)} H_{k-1}^{(1)} }{ H_{k}^{(1)} H_{k}^{(0)} },\quad
  k=1,2,\ldots.
}
From this expression, Hankel determinants $H_m$ can be expressed in a product of $\zeta_k$ or of canonical moments $p_k$,
\onemath{
  H_m^{(0)} &= \prod_{k=1}^{m-1} (\zeta_{2k-1} \zeta_{2k})^{m-k} \\
            &= \left( \prod_{j=1}^{m-1} (1-p_{2j})^{m-j-1} p_{2j}^{m-j} \right) \prod_{j=1}^{m-1} ((1-p_{2j-1}) p_{2j-1})^{m-j},
}{eq:hfoais}
where $2m-2 \leq N$.
D-optimal designs for polynomial regression are calculated in \cite{as33jtuw} through the expression \eqref{eq:hfoais}.
Similar approaches for other D-optimal designs are described in the book \cite{asdfjtuw}.

\section{An algorithm for calculating D-optimal designs for polynomial regression with prior information}

In this section, we consider polynomial regression with prior information.

In Subsection \ref{ss::fiu}, polynomial regression with prior information is defined and formulated as a linear regression.
Then we proposed an algorithm for calculating the D-optimal design for polynomial regression with prior information.

\subsection{Generalized canonical moments and the nonautonomous discrete time Toda equation}

At first, we generalize ordinary moments.
For given moments $\{c_k\}_{k=0}^\infty$, let $c_k^{(T)}$ be defined by
\onemathN{
c_k^{(T \uplus \{\lambda\})} = c_{k+1}^{(T)} - \lambda c_k^{(T)}, \;\; c_k^{(\phi)} = c_k,
}
where $T$ denotes a multiset.
And let $H_k^{(T)}$ be its Hankel determinant $|c_{i+j}^{(T)}|_{i,j=0}^{k-1}$.
Then canonical moments $p_k$ and variables $\zeta_k$ are expressed as
\onemath{
& p_{2k}   = - \frac{H_{k+1}^{(\phi)} H_{k-1}^{(\{0,-1\})}}{ H_k^{(\{0\})} H_k^{(\{-1\})} },
    \;\;
  p_{2k+1} = - \frac{H_{k+1}^{(\{0\})} H_{k}^{(\{-1\})}}{ H_{k+1}^{(\phi)} H_k^{(\{0,-1\})} },
    \\
& \zeta_{2k} = \frac{H_{k+1}^{(\phi)} H_{k-1}^{(\{0\})}}{ H_k^{(\{0\})} H_k^{(\phi)} },
    \;\;
  \zeta_{2k+1} = \frac{H_{k+1}^{(\{0\})} H_{k}^{(\phi)}}{ H_{k+1}^{(\phi)} H_k^{(\{0\})} }.
}{eq;oiad90f}
We define generalized canonical moments $p_k^{(T)}$ and variable $\zeta_k^{(T)}$ by
\onemath{
& p_{2k}^{(T)}   = - \frac{H_{k+1}^{(T)} H_{k-1}^{(T \uplus \{0,-1\})}}{ H_k^{(T \uplus \{0\})} H_k^{(T \uplus \{-1\})} },
    \;\;
  p_{2k+1}^{(T)} = - \frac{H_{k+1}^{(T \uplus \{0\})} H_{k}^{(T \uplus \{-1\})}}{ H_{k+1}^{(T)} H_k^{(T \uplus \{0,-1\})} },
    \\
& \zeta_{2k}^{(T,s)} = \frac{H_{k+1}^{(T)} H_{k-1}^{(T \uplus \{s\})}}{ H_k^{(T \uplus \{s\})} H_k^{(T)} },
    \;\;
  \zeta_{2k+1}^{(T,s)} = \frac{H_{k+1}^{(T \uplus \{s\})} H_{k}^{(T)}}{ H_{k+1}^{(T)} H_k^{(T \uplus \{s\})} }.
}{eq;oiad90g}
It is clear that $p_k^{(\phi)} = p_k, \zeta_k^{(\phi,0)} = \zeta_k$.
While generalized canonical moments do not satisfy like a property \eqref{eq:asfuio7h},
generalized canonical moments satisfy like a property \eqref{eq:iofaou4a}, that is,
\onemath{
  & \zeta_0^{(T)} = 0, \quad \zeta_1^{(T)} = p_1^{(T)}, \\
  & \zeta_k^{(T,0)} = (1-p_{k-1}^{(T)})p_k^{(T)}, \quad k=2,3,\ldots,N. 
}{eq;;as9f}
Additionally variables $\zeta_k^{(T)}$ is a determinant solution of nonautonomous discrete time Toda equation, 
namely, $\zeta_k^{(T)}$ is satisfy nonautonomous discrete time Toda equation
\onemath{
& \zeta_{2k}^{(T \uplus \{\lambda_1\}, \lambda_2)} + \zeta_{2k+1}^{(T \uplus \{\lambda_1\}, \lambda_2)} + \lambda_2 = \zeta_{2k+1}^{(T,\lambda_1)} + \zeta_{2k+2}^{(T,\lambda_1)} + \lambda_1, \\
& \zeta_{2k+1}^{(T \uplus \{\lambda_1\}, \lambda_2)} \zeta_{2k+2}^{(T \uplus \{\lambda_1\}, \lambda_2)} = \zeta_{2k+2}^{(T,\lambda_1)} \zeta_{2k+3}^{(T,\lambda_1)}.
}{eq:asiu9aToda}
From \eqref{eq:asiu9aToda}, we obtain the following formula
\onemath{
  & \zeta_{2k}^{(T, \lambda_1)} + \zeta_{2k+1}^{(T, \lambda_1)} + \lambda_1 = \zeta_{2k+1}^{(T,\lambda_2)} + \zeta_{2k+2}^{(T,\lambda_2)} + \lambda_2, \\
  & \zeta_{2k+1}^{(T, \lambda_1)} \zeta_{2k+2}^{(T, \lambda_1)} = \zeta_{2k+2}^{(T,\lambda_2)} \zeta_{2k+3}^{(T,\lambda_2)}.
}{eq:dafopfa90}

\subsection{D-optimal designs for polynomial regression with prior information} \label{ss::fiu}

We consider the $(m+S-1)$-th degree of polynomial regression
\onemath{
  & Y = \sum_{k=0}^{m+S-1} \theta_k x^k + \varepsilon,\\
  & E[\varepsilon] = 0, \quad V[\varepsilon] = \sigma^2
}{eq:oiaufoa9}
with the $S = \sum_{j=1}^l b_j$ values as prior information
\onemath{
\left. \frac{\diffd^k}{\diffd x^k} g(x) \right|_{x=\beta_j}, \quad 0 \leq j < l, \quad 0 \leq k < b_j,
}{eq:oripr}
where $g(x) = E[Y|x] = \sum_{k=0}^{m+S-1} \theta_k x^k$, 
$b_0, b_1, \ldots, b_{l-1}$ are positive integers,
and $\beta_0, \beta_1, \ldots, \beta_{l-1}$ are arbitrary distinct real values.
That means we consider the cases where the exact $S$ values \eqref{eq:oripr} are known before experiments.
Note that $\beta_k$ do not have to be in $\mathcal X = [0,1]$.
We call the model \eqref{eq:oiaufoa9} ${\rm PRM}_m(\beta, b)$, for short, where $\beta = (\beta_0, \beta_1, \ldots, \beta_{l-1})$, $b = (b_0, b_1, \ldots, b_{l-1})$.

There are multiple ways to consider the model ${\rm PRM}_m(\beta, b)$ as linear regression models.
However the D-optimal design for the model ${\rm PRM}_m(\beta, b)$ is defined uniquely,
since D-optimal designs for linear regression models depend on a linear space spanned by bases functions only (see \cite[Theorem 5.5.1]{asdfjtuw}).
The D-optimal designs are formulated as the following theorem.
A proof of the theorem is given in Appendix.
\begin{theo} \label{theo:a89fas}
The D-optimal design for polynomial regression with prior information ${\rm PRM}_m(\beta, b)$ is defined as the optimal solution of the optimization problem
\onemath{
& \mbox{maximize}\; H_m^{(T)} \\
& \mbox{subject to}\; \mu \in \mathcal P_{[0,1]}.
}{oofdifufyu}
where the multiset $T$ satisfies the number of $\beta_k$ in the $T$ is $2 b_k$, that is, let $m_T(x)$ be the multiplicity function, then
\onemath{
m_T(x) = \begin{cases}
2b_k & (x=\beta_k) \\
0 & (\mbox{\rm otherwise})
\end{cases}.
}{eq:foiuefaoT}
\end{theo}
Note that the D-optimal design for ${\rm PRM}_m(\beta, b)$ is equivalent to the D-optimal design for 
weighted polynomial regression model
\onemathN{
& Y = \sum_{k=0}^{m-1} \theta_k x^k + \varepsilon,\\
& E[\varepsilon] = 0, \quad V[\varepsilon] = \frac{\sigma^2}{w(x)}, \quad w(x) = \prod_{k=0}^{l-1} (x-\beta_k)^{2b_k}.
}

Here we propose an algorithm for rephrasing the optimization problem \eqref{oofdifufyu} in terms of canonical moments.
In the algorithm, the two formulas \eqref{eq:afa908}, \eqref{eq:a90dfadk} and the nonautonomous discrete time Toda equation \eqref{eq:asiu9aToda} are used.

At first, we describe the objective function $H_m^{(T)}$ by using the value $\zeta_k^{(T,s)}$ corresponding to generalized canonical moments.
We obtain the formula which is similar to \eqref{eq:hfoais}
\onemath{
  H_m^{(T)} = \left( c_0^{(T)} \right)^m \prod_{j=1}^{m-1} \left( \zeta_{2j-1}^{(T,0)} \zeta_{2j}^{(T,0)} \right)^{m-j}.
}{eq:afa908}

Then we describe the objective function $H_m^{(T)}$ in terms of the value $\zeta_k^{(\phi,0)}$ corresponding to canonical moments.
Here by using the nonautonomous discrete time Toda equation \eqref{eq:asiu9aToda} and the formula \eqref{eq:dafopfa90},
$\zeta_k^{(T,s)}$ can be expressed in terms of $\zeta_1^{(\phi,0)}, \zeta_2^{(\phi,0)}, \ldots$.
And we obtain the formula about $c_0^{(T)}$
\onemath{
  \zeta_1^{(T,s)} = \frac{c_0^{(T \uplus \{s\})}}{c_0^{(T)}}, \quad c_0^{(\phi)} = 1,
}{eq:a90dfadk}
then $c_0^{(T)}$ also can be expressed in terms of $\zeta_1^{(\phi,0)}, \zeta_2^{(\phi,0)}, \ldots$ by using the formula.

Lastly we describe the objective function $H_m^{(T)}$ in terms of canonical moments by using the relationship \eqref{eq:iofaou4a}.

By putting it all together, 
the proposed algorithm for calculating the D-optimal design for polynomial regression with prior information \eqref{eq:oiaufoa9} 
is described as follows.
\noindent
\begin{center}

\framebox[12cm][c]{
\begin{minipage}{11.5cm}

{\bf The algorithm for calculating D-optimal designs for ${\rm PRM}_m(\beta, b)$}
\begin{enumerate}[Step 1.] \setlength{\parskip}{0pt} \setlength{\itemsep}{0pt}
\item
By using the formula \eqref{eq:afa908},
describe the objective function $H_{m}^{(T)}$ in terms of $\zeta_k^{(T,s)}$ and $c_0^{(T)}$.
\item
By using the nonautonomous discrete time Toda equation \eqref{eq:asiu9aToda} and the formulas \eqref{eq:dafopfa90}, \eqref{eq:a90dfadk},
describe the objective function $H_{m}^{(T)}$ in terms of $\zeta_k^{(\phi,0)} = \zeta_k$.
\item
By using the relationship \eqref{eq:iofaou4a},
describe the objective function $H_{m}^{(T)}$ in terms of canonical moments $p_k$.
\item
Find canonical moments which maximize the objective function $H_{m}^{(T)}$.
\end{enumerate}

\end{minipage}
}
\end{center}

\section{Application of the algorithm for calculating some D-optimal designs}

In this section, we introduce two examples of application of our algorithm.
In Subsection \ref{sss:df98dss}, robust D-optimal designs for approximate polynomial regression with prior information is considered.
This model is generalized model considered by \cite{asfa831221i}.
In Subsection \ref{sss:asd9f9ai}, maximin optimal designs for estimating a function of parameters on weighted polynomial regression.
We consider the generalized case of the case considered by \cite{ajifa9d78rjio}.

\subsection{Robust D-optimal designs for approximate polynomial regression with prior information} \label{sss:df98dss}

In this subsection, let the design space be $\mathcal X = [-1,1]$ instead of $[0,1]$.
There is one to one correspondence between $\mu \in \mathcal P_{[0,1]}$ and a symmetric measure $\xi \in \mathcal P_{[-1,1]}$
such that 
\onemathN{
   \mu([0,x^2]) = \xi([-x,x]).
}

The approximate polynomial regression model is described as
\onemathN{
  & Y = \sum_{k=0}^{m-1} \theta_k x^k + x^m \psi(x) + \varepsilon, \\
  & E[\varepsilon] = 0, \quad V[\varepsilon] = \sigma^2,
}
where $\psi$ denotes an unknown function.
Let the best linear unbiased estimator be $\hat \theta (\psi)$ when we estimate as $\psi(x)$ is considered as 0.
Then we obtain
\onemathN{
  & E[\hat \theta (\psi) - \hat \theta (0)] = B_m^{(\phi)}(\xi) r(\phi) \\
  & V[\hat \theta (\psi)] = (\sigma^2 / n) B_m^{(\phi)}(\xi)
}
where $\xi$ is a design, $n$ is the number of observations, and
\onemathN{
  & r(\psi) = \int_{-1}^1 (1,x,\ldots,x^{m-1})^{\rm T} x^m \psi(x) \diffd \xi(x),\\
  & B_m^{(\phi)} = (c_{i+j}(\xi))_{i,j=0}^{m-1}.
}
For given a continuous function $\eta(x)$ and a positive number $d$, maximin optimal design is defined as
\onemathN{
  & \mbox{maximize}\; H_m^{(\phi)}(\xi) \\
  & \mbox{subject to}\; \xi \in \mathcal P_{[-1,1]}, \;\; \sup_{|\psi| \leq |\eta|}r(\psi)^{\rm T} (B_m^{(\phi)})^{-1} r(\psi) \leq d.
}
In \cite{asfa831221i}, the case where 
\onemath{
  \eta(x) = |x|^\alpha, \;\; \alpha \in {\mathbb Z}_{\geq 0}
}{eq:aioufa9f8}
is considered, and it is shown that the constraint $\sup_{|\psi| \leq |\eta|} r(\psi)^{\rm T} (B_m^{(\phi)})^{-1} r(\psi) \leq d$ is described in terms of canonical moments as
\onemath{
  \sup_{|\psi| \leq |\eta|}r(\psi)^{\rm T} (B_m^{(\phi)})^{-1} r(\psi) = \sum_{i = \lfloor \alpha / 2 \rfloor + 1}^{\lfloor (m+\alpha)/2 \rfloor} S_{i, m+\alpha-i}(\mu)^2 \prod_{j=1}^{m+\alpha-2i} \zeta_j (\mu),
}{eq:aiofa908a}
where $S_{i,j}(\mu)$ is defined recursively as 
\onemathN{
  & S_{i,j}(\mu) = 0, \quad 0 \leq j < i, \\
  & S_{i,j}(\mu) = 1, \quad i=0, \;\; j > 0, \\
  & S_{i,j}(\mu) = S_{i,j-1}(\mu) + \zeta_{j-i+1}(\mu) S_{i-1,j}(\mu), \quad 0 < i \leq j.
}

Now we consider the approximate polynomial regression model with prior information, namely,
\onemath{
  & Y = \sum_{k=0}^{2S+m-1} \theta_k x^k + x^{2S+m} \psi(x) + \varepsilon, \\
  & E[\varepsilon] = 0, \quad V[\varepsilon] = \sigma^2,
}{eq:apo90af8f}
with the $2S$ values as prior information
\onemath{
& \left. \frac{\diffd^k}{\diffd x^k} g(x) \right|_{x=\beta_j},\\
& \left. \frac{\diffd^k}{\diffd x^k} g(x) \right|_{x=-\beta_j}, \quad 0 \leq j < l, \quad 0 \leq k < b_j,
}{eq:sdapufa90}
where $g(x) = E[Y|x] = \sum_{k=0}^{m+S-1} \theta_k x^k$, and $\beta_0, \beta_1, \ldots, \beta_{l-1}$ are arbitrary distinct real values.
Note that prior information \eqref{eq:sdapufa90} must be symmetric with respect to the origin.

We can show that the optimization problem corresponding to the model \eqref{eq:apo90af8f} is the following by similar calculation to \cite{asfa831221i}
\onemath{
  & \mbox{maximize}\; H_m^{(T')}(\xi) \\
  & \mbox{subject to}\; \xi \in \mathcal P_{[-1,1]}, \;\; \sup_{|\psi| \leq |\eta|}r^{(T')}(\psi)^{\rm T} (B_m^{(T')})^{-1} r^{(T')}(\psi) \leq d,
}{auio3pfau}
where $T'$ denotes the multiset satisfying
\onemathN{
  m_{T'}(x) = \begin{cases}
    2b_k & (x=\beta_k) \\
    2b_k & (x=-\beta_k) \\
    0 & (\mbox{\rm otherwise})
  \end{cases},
}
and
\onemathN{
  & B^{(T{}')}_m (\xi) = (c_{i+j}^{(T{}')})_{i,j=0}^{m-1} \\
  & r^{(T')} (\psi) = \int_{-1}^1 (1,x,\ldots,x^{m-1})^{\rm T} x^m \psi(x) \left( \prod_{j=0}^{l-1} (x-\beta_j)(x+\beta_j) \right) \diffd \xi(x).
}

The constraint of \eqref{auio3pfau} corresponding to \eqref{eq:aiofa908a} is described in terms of generalized canonical moments as
\onemathN{
  (c_0^{(T')}(\xi))^{-1} \sum_{i = \lfloor \alpha / 2 \rfloor + 1}^{\lfloor (m+\alpha)/2 \rfloor} S_{i, m+\alpha-i}^{(T)}(\mu)^2 \prod_{j=1}^{m+\alpha-2i} \zeta_j^{(T,0)} (\mu),
}
where $T$ is the same as \eqref{eq:foiuefaoT}, and
\onemathN{
  & S_{i,j}^{(T)}(\mu) = 0, \quad 0 \leq j < i, \\
  & S_{i,j}^{(T)}(\mu) = 1, \quad i=0, \;\; j > 0, \\
  & S_{i,j}^{(T)}(\mu) = S_{i,j-1}^{(T)}(\mu) + \zeta_{j-i+1}^{(T,0)}(\mu) S_{i-1,j}^{(T)}(\mu), \quad 0 < i \leq j.
}
Hence we can obtain the expression of the optimization problem corresponding \eqref{eq:apo90af8f} in terms of canonical moments by using our algorithm.

\subsection{Maximin optimal designs for estimating a function of parameters on weighted polynomial regression} \label{sss:asd9f9ai}

Consider the weighted polynomial regression
\onemathN{
  & Y = \sum_{k=0}^{m-1} \theta_k x^k + \varepsilon, \\
& E[\varepsilon] = 0, \quad V[\varepsilon] = \sigma^2 w(x), \quad w(x) = \prod_{k=0}^{l-1} (x-\beta_k)^{2b_k}.
}
In this subsection, we consider an optimal design for estimating $g_{m-1}(\theta_{m-1}) + g_{m-2}(\theta_{m-2}) + \cdots + g_{0}(\theta_{0})$,
where $g_k$ is a polynomial.
In this case, the inverse of the asymptotic variance of the estimator is
\onemathN{
  \gamma(\mu,\theta) = \sum_{k=0}^{m-1} \left( \frac{{\rm d}}{{\rm d}\theta_{k}}g_k(\theta_{k}) \right)^2 \psi_{k}^{(1)}(\mu),
}
where $\psi_k^{(1)}(\mu) = H_{k+1}^{(T)} / H_{k}^{(T)}$, and $T$ is the same as \eqref{eq:foiuefaoT}.
Since the variance of estimator depends on the unknown parameters $\theta_k$, we consider
maximin optimal design defined as the optimization solution of the optimization problem
\onemathN{
& \mbox{maximize}\; \min_{\theta \in \Theta} \gamma(\mu,\theta) \\
& \mbox{subject to}\; \mu \in \mathcal P_{[0,1]}
}
where $\Theta$ is a given parameter space.
Suppose the parameter space $\Theta = \{\theta \;|\; s_k \leq \theta_k \leq t_k \}$.

From \cite[Theorem 3.1]{ajifa9d78rjio}, the optimal solution of the optimization problem
\onemath{
  & \mbox{maximize}\; \int_{\Theta} \gamma(\mu,\theta)^p {\rm d}\pi (\theta) \\
  & \mbox{subject to}\; \mu \in \mathcal P_{[0,1]}.
}{eq:aospfa908}
converges weakly to the maximin optimal design as $p \to -\infty$, where
\onemathN{
  & {\rm d}\pi(\theta) = {\rm d}\theta \prod_{k=0}^{m-1} h_k(\theta_k), \\
  & h_k(\theta_k) = \begin{cases}
      \displaystyle \frac{{\rm d}}{{\rm d}\theta_{k}} \left( \frac{{\rm d}}{{\rm d}\theta_{k}}g_k(\theta_{k}) \right)^2 & (\deg g_k \geq 2) \\
      \displaystyle 1 & (\mbox{otherwise})
    \end{cases}.
}
Then we can calculate the integral in the objective function of the \eqref{eq:aospfa908}.
And we can express $\psi_k^{(1)}(\mu)$ in terms of canonical moments by our algorithm.
Therefore, after expressing the optimization problem \eqref{eq:aospfa908} in terms of canonical moments,
we can obtain an approximate maximin optimal design by solving \eqref{eq:aospfa908} numerically for small $p$.

\appendix

\section{The proof of Theorem \ref{theo:a89fas}}

It can be turn out that the optimization problem \eqref{oofdifufyu} corresponding to ${\rm PRM}_m(\beta, b)$ corresponds to the vector of basis functions
\onemath{
  f(x) = \left( \prod_{j=0}^{l-1} (x - \beta_j)^{b_j} \right) (1, x, \ldots, x^{m-1})^{\rm T}.
}{eqjidioafy}
Let $g_k(x) = (1, x, \ldots, x^{k-1})$ be the vector of basis functions for polynomial regression,
then the vector \eqref{eqjidioafy} of basis functions corresponding to ${\rm PRM}_m((\beta_0,\beta_1,\ldots,\beta_{l-1})$, $(b_0,b_1,\ldots,b_{l-1}))$ is expressed as
\onemath{
  f(x) = \prod_{j=0}^{l-1} (x - \beta_j)^{b_j} g_m(x).
}{eqdiduad}

To prove Theorem \ref{theo:a89fas}, it suffices to show that ${\rm PRM}_m((\beta_0,\beta_1,\ldots,\beta_{l-1})$, $(b_0,b_1,\ldots,b_{l-1}))$ corresponds to the vector \eqref{eqdiduad} of basis functions.
Let us prove it by the principle of induction.
By the symmetricity of $\beta_j$ and $b_j$, we can assume that 
${\rm PRM}_{m+1}((\beta_0,\beta_1,\ldots,\beta_{l-1})$, $(b_0,b_1-1,\ldots,b_{l-1}))$ corresponds to the vector of basis functions
\onemathN{
  f(x) = (x - \beta_0)^{b_0-1} \left( \prod_{j=1}^{l-1} (x - \beta_j)^{b_j} \right) g_{m+1}(x).
}
Let $M(x) = \prod_{j=1}^{l-1} (x - \beta_j)^{b_j}$, then the linear regression model ${\rm PRM}_{m+1}((\beta_0,\beta_1,\ldots,\beta_{l-1})$, $(b_0-1,b_1,\ldots,b_{l-1}))$
is described as
\onemathN{
  Y = (x - \beta_0)^{b_0-1} M(x) \sum_{k=0}^{m} \theta_k x^k + \varepsilon.
}
Thus the linear regression model ${\rm PRM}_m((\beta_0,\beta_1,\ldots,\beta_{l-1})$, $(b_0,b_1,\ldots,b_{l-1}))$ is described as
\onemath{
  Y = (x - \beta_0)^{b_0-1} M(x) \sum_{k=0}^{m} \theta_k x^k + \varepsilon.
}{aioufoaidfa}
with one given value as prior information
\onemath{
  \left. \frac{\diffd^{b_0-1}}{\diffd x^{b_0-1}} \left( (x - \beta_0)^{b_0-1} M(x) \sum_{k=0}^{m} \theta_k x^k \right) \right|_{x=\beta_0}.
}{auifdfa98ua}
Since
\onemathN{
  \left. \frac{\diffd^k}{\diffd x^k} (x - \beta_0)^{b_0-1}  \right|_{x=\beta_0} = 0, \quad k = 0, 1, \ldots, b_0-2,
}
the value \eqref{auifdfa98ua} becomes
\onemathN{
  (b_0 -1)! M(\beta_0) \sum_{k=0}^m \theta_k \beta_0^k
}
by the general Leibniz rule.
Hence we obtain the value
\onemath{
  \alpha = \sum_{k=0}^m \theta_k \beta_0^k
}{eq:iaufopia}
from prior information \eqref{auifdfa98ua}.

From \eqref{eq:iaufopia}, substitute $\theta_0 = \alpha - \sum_{k=1}^m \theta_k \beta_0^k$ into the model \eqref{aioufoaidfa},
we obtain
\onemath{
  & Y = (x - \beta_0)^{b_0-1} M(x) \left( \sum_{k=1}^{m} \theta_k x^k - \sum_{k=1}^m \theta_k \beta_0^k + \alpha \right) + \varepsilon.
}{eq:afliauf9}
When we obtain a response $y_k$ by the obserbation at the experimental condition $x_k$,
we can calculate the value $y_k - (x_k-\beta_0)^{b_0-1} M(x_k) \alpha$ easily.
Thus we can ignore the term $(x-\beta_0)^{b_0-1} M(x) \alpha$ from the model \eqref{eq:afliauf9}, then we obtain the model
\onemath{
  & Y = (x - \beta_0)^{b_0-1} M(x) \sum_{k=1}^{m} \theta_k (x^k - \beta_0^k) + \varepsilon.
}{eq:adfsaauf9}
Here the vector of basis functions corresponding to the model \eqref{eq:adfsaauf9} is
\onemath{
  f(x) = (x - \beta_0)^{b_0-1} M(x)
  \begin{pmatrix}
    x - \beta_0 \\
    x^2 - \beta_0^2 \\
    \vdots \\
    x^m - \beta_0^m
  \end{pmatrix}.
}{eq:adfpoaipfo}
Let the non-singular matrix $A$ be
\onemathN{
  A = \begin{pmatrix}
    1 & 0 & 0 & \cdots & 0 \\
    -\beta_0 & 1 & 0 & \cdots & 0 \\
    0 & -\beta_0 & 1 & \cdots & 0 \\
    \vdots & \vdots & \vdots & \ddots & \vdots \\
    0 & 0 & 0 & \cdots &  1 \\
  \end{pmatrix}
}
then, by multiplying $A$ to the vector \eqref{eq:adfpoaipfo} of basis functions from left,
\onemathN{
  Af(x) 
  &= (x - \beta_0)^{b_0-1} M(x) 
   \begin{pmatrix}
     x - \beta_0 \\
     x^2 - \beta_0^2 - \beta_0 (x - \beta_0) \\
     \vdots \\
     x^m - \beta_0^m - \beta_0 (x^{m-1} - \beta_0^{m-1})
   \end{pmatrix} \\
  &= (x - \beta_0)^{b_0-1} M(x) 
   \begin{pmatrix}
     x - \beta_0 \\
     x (x - \beta_0) \\
     \vdots \\
     x^{m-1} (x - \beta_0)
   \end{pmatrix} \\
  &= (x - \beta_0)^{b_0} M(x) g_m(x).
}
Thus we obtain \eqref{eqdiduad}.


\bibliographystyle{main}
\bibliography{main}

\end{document}